# Improvements in Google Scholar Citations are for the summer: creating an institutional affiliation link feature

Enrique Orduna-Malea[1], Juan Manuel Ayllón[2], Alberto Martín-Martín[2], Emilio Delgado López-Cózar[2]

[1] EC3: Evaluación de la Ciencia y de la Comunicación Científica, Universidad Politécnica de Valencia (Spain)
[2] EC3: Evaluación de la Ciencia y de la Comunicación Científica, Universidad de Granada (Spain)

**ABSTRACT**

This report describes the feature introduced by Google to provide standardized access to institutional affiliations within Google Scholar Citations. First, this new tool is described, pointing out its main characteristics and functioning. Next, the coverage and precision of the tool are evaluated. Two special cases (Google Inc. and Spanish Universities) are briefly treated with the purpose of illustrating some aspects about the accuracy of the tool for the task of gathering authors within their appropriate institution. Finally, some inconsistencies, errors and malfunctioning are identified, categorized and described. The report finishes by providing some suggestions to improve the feature. The general conclusion is that the standardized institutional affiliation link provided by Google Scholar Citations, despite working pretty well for a large number of institutions (especially Anglo-Saxon universities) still has a number of shortcomings and pitfalls which need to be addressed in order to make this authority control tool fully useful worldwide, both for searching purposes and for metric tasks.

# 1. INTRODUCTION

It seems that the Mountain View's company has a special fondness for the summer to make changes to its flagship products. If last year it announced on August 21st a "_Fresh Look of Scholar Profiles_"[1], this 2015 we have learnt almost at the same time - not from the official Google Scholar blog, which has not provided any information, but from a Tweet by Isidro Aguillo[2] - that "_Google Scholar Citations add links to institution`s names (incl acronyms) in correct-built affiliations of profiles_".

Indeed, the degree of control in Google Scholar's searches was a topic openly discussed during the _15th International Conference on Scientometrics & Informetrics_ (ISSI-2015), specifically in the workshop "Google Scholar and related products", in which the _EC3 Research Group_ had the opportunity to participate. Interestingly, _Isidro Aguilló_ emphasized the need for more controlled tools to improve the achievement of accurate results that could be better used in bibliometric tasks, such as using ORCID for author identification or results filtered by institution, among other recommendations.

We partially discussed the feasibility of these ideas considering the traditional point of view of an academic search engine (and a company) that has always been characterized by natural language searches and giving full freedom to the user in order to lose some terminological control in recovery. We must not forget that _Google Scholar_ has not been primarily developed for bibliometric purposes.

However, breaking a little from its usual approach, _Google Scholar Citations_ has implemented a new information search feature under the form of an authority control tool for institutional affiliations, which lies halfway between the classic controlled search and the natural language, the comprehensiveness and precision of which will be evaluated in this Report.

We definitely welcome this new initiative, which represents an improvement in the product since it allows having a new and easy way to search information from scholars belonging to a specific institution. Previously, specific searches by the institution name or the email domain in the open box were required for this, a tedious and very unfriendly process.

Now, just by clicking on the name of the institution we can identify all scholars belonging to an organization, as well as the global scientific interest and thematic focus of the corresponding institution. Incidentally, it will facilitate the morbid – as well as dangerous - evaluative exercises that some institutions have already performed based on the data available in this platform.

## 2. GENERAL FUNCTIONING

In Figure 1 we can observe an example of this new tool, available in different language interfaces and not only in the main English one. In the area reserved for the professional affiliation we can see how the institution appears as a hyperlink in the same way as thematic keywords appear. By clicking on the hyperlink the system returns a list of authors affiliated to the corresponding institution (see Figure 1; down). With this simple and direct procedure, the user can navigate within an institution, in which the authors are sorted in descending order according to the total number of citations received.

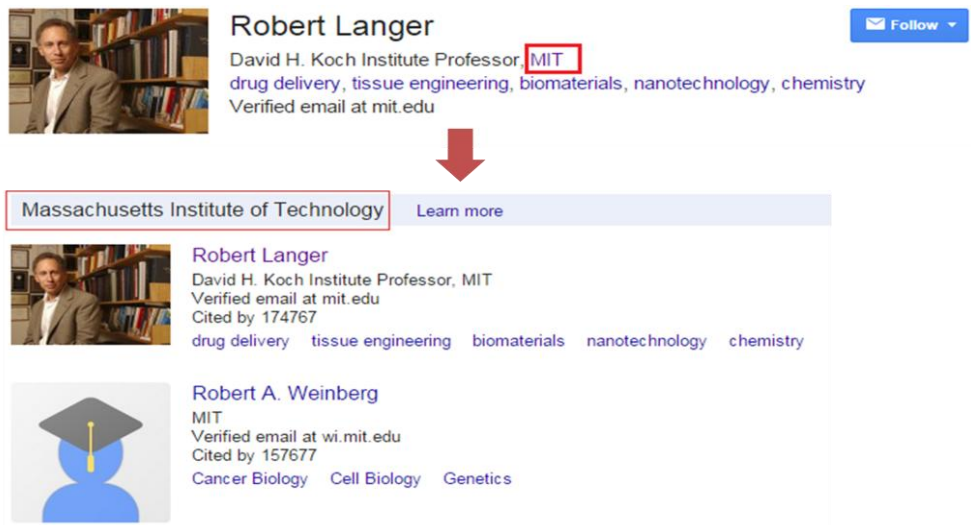

**Fig 1. Linked standardized institutional affiliations in *Google Scholar Citations***

The tool does not allow direct search for institutions from the general *Google Scholar's* search box, but only from the internal search box provided by *Google Scholar Citations*. Thus, if we make a search for a particular keyword (e.g. "Stanford"), the system returns a set of authors in which the word "*Stanford*" appears in the bibliographic description, but also identifies institutions that contain that word (e.g. "*Stanford University*"), thus allowing direct access to the authors of that institution (Figure 2):

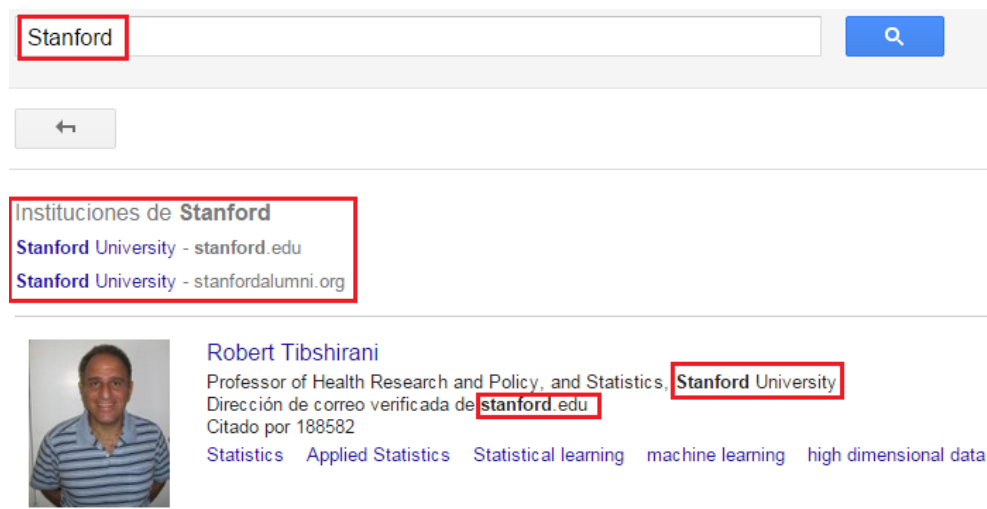

**Fig.2. Institutional search by keyword in Google Scholar Citations**

Perhaps one of the main virtues of this new feature introduced by *Google* is the localization and recognition of the different variants of an institution's name in different languages, and selecting as the primary entry the name that corresponds with the vernacular language.

For example, if an author has identified its Spanish academic institution under its nomenclature in other languages (*Universidad de Barcelona*, *University of Barcelona* or even *Barcelona University*), the tool redirects the user to the "*Universitat de Barcelona*" entry, which is the official name of the university (Figure 3). Moreover, the acronym of the institution (UB) is considered as a variant itself (UB). That is, authority control works automatically for institutional variants in different languages.

**Fig.3. Authority control for institutional name variants in different languages**

Furthermore, the feature is capable of distinguishing institutions that unfortunately use the same institutional acronym but different URLs, such as *Universitat Politècnica de València* and *Euskal Herriko Unibertsitatea* (Figure 4). Despite the acronym for Euskal Herriko Unibertsitatea is EHU, it is commonly used UPV, which corresponds with the official acronym for Universitat Politècnica de València, generating confussion.

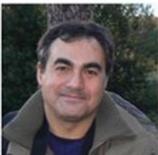

**Fig.4. Authority control for identical acronyms corresponding to different institutions**

# 3. AUTHORITY CONTROL

Those who have been dedicated to bibliographic work for years and, of course, librarians, who have been building catalogs for centuries, allowing access to documents stored in libraries (Figure 5), know how hard it is to get a proper identification and unification for the names of things, whether they be individuals, organizations, places or topics.

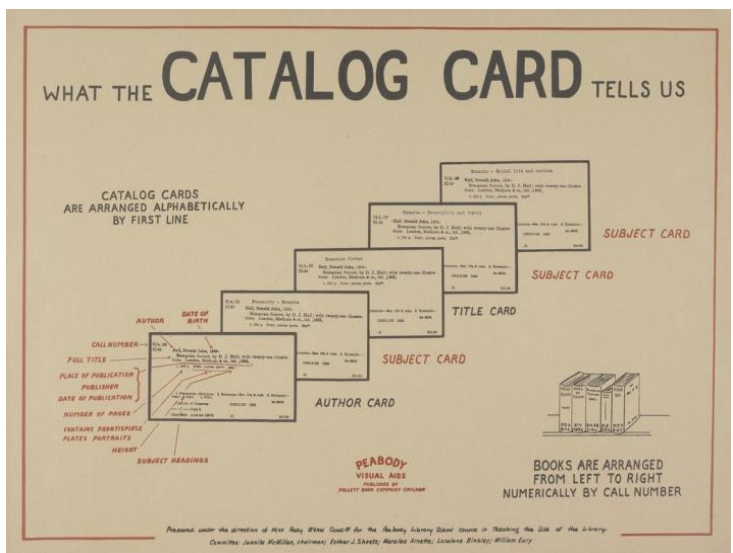

**Figure 5a: Example of the functioning of an old printed catalog card**
Source: Flickr by Char Booth (informational.com)

Authority control[3] is the soul of our venerable profession and the process that aims to achieve - through various initiatives (such as the *Virtual International Authority File*[4]) – a universal bibliographic control that will generate uniform titles.

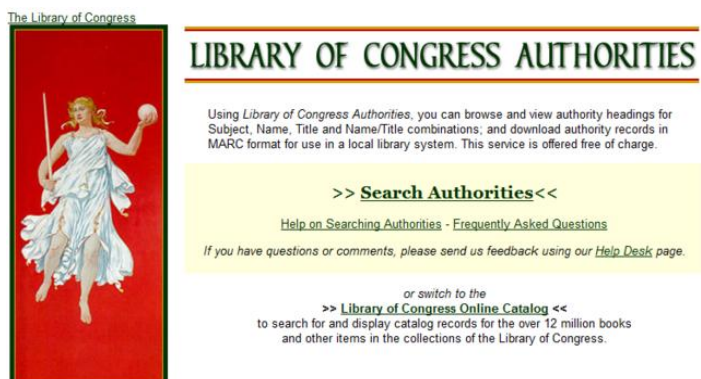

**Figure 5b: Home page Library of Congress Authorities**
Source: http://authorities.loc.gov

The bibliographic reality is multifaceted and full of unexpected ocurrences. Authors are capricious and inconsistent by nature in providing their professional affiliations (from using the most diverse non-standardized variants to merely forgetting to indicate it). When gathering all members of an institution, the number of different denominations may become endless.

EC3 Reports 14In Figure 6 we can observe the different denominations for "*Universidad de Granada*" and internal entities (departments, faculties) taken from *Science Citation Index and Medline* (1987-1993).

**Tabla I.** Denominaciones de entidades de la Universidad de Granada en las bases de datos *Science Citation Index* y *Medline* (1987-1993).

| Denominaciones entidades | SCI | Medline | Denominaciones entidades | SCI | Medline |
|---|---|---|---|---|---|
| UNIVERSidad de Granada | | | University of Granada School of Medicine | 7 | – |
| Granada Univ | – | 3 | University School of Medicine | 2 | – |
| Granada-Univ | – | 3 | University School of Medicine and Dentistry | 1 | – |
| Granada University | 4 | – | | | |
| Unive | 4 | – | DEPARTAMENTO DE MEDICINA LEGAL Y PSIQUIATRÍA | | |
| Univ-Granada | – | 275 | Cátedra de Medicina Legal | 7 | – |
| Univ Granada | – | 1 | Cátedra de Medicina Legal y Toxicología | 1 | – |
| University of Granada | 1 | – | Departament de Medicina Legal | 1 | – |
| Universidad de | 3 | – | Departamento de Medicina Legal | 2 | – |
| Universidad de Gra | 1 | – | Departamento de Medicina Legal y Toxicología | 1 | – |
| Universidad de Grana | 2 | – | Department of Legal Medicine | 7 | – |
| Universidad de Granada | 155 | – | Department of Legal Medicine and Toxicology Service | 3 | – |
| Universit | 1 | – | Dept Legal Med | – | 10 |
| Universitat Granada | 1 | – | Dept Legal Med & Toxicol | – | 1 |
| University Granada | 2 | – | Dept Legal Med & Toxicol Serv | – | 1 |
| University of | 2 | – | Dept Med Legal | – | 6 |
| University of Granada | 197 | – | Dept Med Legal 6 Toxicol | – | 1 |
| Univesity of Granada | 1 | – | Legal Med | 1 | – |
| FACULTAD DE MEDICINA | 1 | – | DIRECCIÓN POSTAL | | |
| Facultad de Medicina | 99 | – | Av Madrid 11 | 2 | – |
| Fac Med Granada | – | 127 | Av Madrid 9 | 1 | – |
| Fac Med & Dent | – | 3 | Av Madrid s-n | 3 | – |
| Fac Med & Odon | – | 1 | Av Madrid sn | 1 | – |
| Fac Med & Pharm | – | 2 | Avd Madrid | 1 | – |
| Fac Med Granada | – | 24 | Avd Madrid 11 | 4 | – |
| Fac-Med-Granada | – | 18 | Avd Madrid s-n | 1 | – |
| Facultat de Medicina | 1 | – | Avda Madrid | 2 | – |
| Facultad de Medi | 1 | – | Avda Madrid 11 | 15 | – |
| Faculty of Medicine | 40 | – | Avda Madrid 12 | 2 | – |
| Faculty of Medicine and Dentistry | 2 | – | Avda Madrid 9 | 3 | – |
| Medical Faculty of Granada | 1 | – | Avda Madrid n-s | 1 | – |
| Med Sch Granada | – | 1 | Avda Madrid s n | 1 | – |
| Medical School | 6 | – | Avda Madrid s-n | 21 | – |
| Medizinischen Fakultat | 7 | – | Avda Madrid sn | 2 | – |
| Sch Dent Med | – | 1 | Avda Madrid | 1 | – |
| Sch Med | – | 90 | Ave Madrid | 3 | – |
| Sch Med & Dent | – | 1 | Avda Madrid 11 | 5 | – |
| Sch Med Granada | – | 1 | Avda Madrid 9 | 1 | – |
| School of Medicine | 43 | – | Avda Madrid s-n | 3 | – |
| School of Medicine and Dentistry | 1 | – | | | |
| University of Granada Medical School | – | 7 | | | |

**Fig.6.** Institutional variants for *Universidad de Granada* (**SSC and Medline**) [5]

## 4. OBJECTIVES

Constituting this the stubborn reality, we must ask the following questions about the new tool developed by *Google Scholar Citations* to achieve a correct identification of institutional affiliations for scholars:

- How has a technological giant like *Google* solved this problem?
- Has *Google Scholar Citations* been able to identify all institutions mentioned in the profiles?
- Has *Google Scholar Citations* been able to create a unique preferred name representing all variants, including different spellings and misspellings, uppercase and lowercase, acronyms and abbreviations?
- Has *Google Scholar Citations* been able to correctly ascribe all scholars to the organization they're associated with?

All screenshots, examples, data, and analysis have been taken and performed on September 8[th] 2015.

## 5. RESULTS

We must alert users that *Google Scholar Citations* hasn't got a magic wand to magically solve the problem of authority control. It has done a good and meritorious work but is still far from having achieved a fully satisfactory institutional control for all the authors in its system.

Next we give an account of each of the pitfalls we found about the functioning of the institutional authority control feature in *Google Scholar Citations*, with the hope that this may be helpful to identify and solve some errors that could help improve the product.

### *Coverage: are all institutions standardized?*

To begin with, we must highlight that *Google Scholar Citations* has been able to identify many of the organizations, but not all of them. To cite but one example, the following randomly chosen universities: *Universidad Antonio de Nebrija, Universidad Nacional de Loja, Universidad del Valle de México, Pontificia Universidad Católica Argentina, Universidade Católica de Pelotas, Universidad Católica de Ávila, Universidad Católica de Santiago de Guayaquil*.

Of course, these are not world-class universities, but this surfaces a problem in the identification of institutions appearing in author profiles, which makes this feature still not fully satisfactory.

To calibrate the volume of this problem we would have to conduct a comprehensive study. However, for the mere goal of illustrating this problem, we have employed as a small sample the 82 existing Spanish universities today (September 2015), obtaining that 13 of them (15.85%) are not listed with a standardized *Google Scholar Citations* link.

### *Precision: are all authors catalogued in the corresponding institution?*

Perhaps the main problem encountered is that *Google Scholar Citations* has failed to gather all profiled scholars who work or are affiliated with a certain institution. The reason for this falls in the criterion used for the "author-organization" link: only those authors who have simultaneously indicated the name of the institution in its affiliation and have verified the profile using an e-mail address from that same institution are properly included in the list of researchers of that institution.

To illustrate this, nothing better than showing the profiles of the two fathers of *Google Scholar Citations* (Figure 7). As we can see, *Anurag Acharya's* profile is correctly included, while the profile of *Alex Verstak* is not linked to any institution.

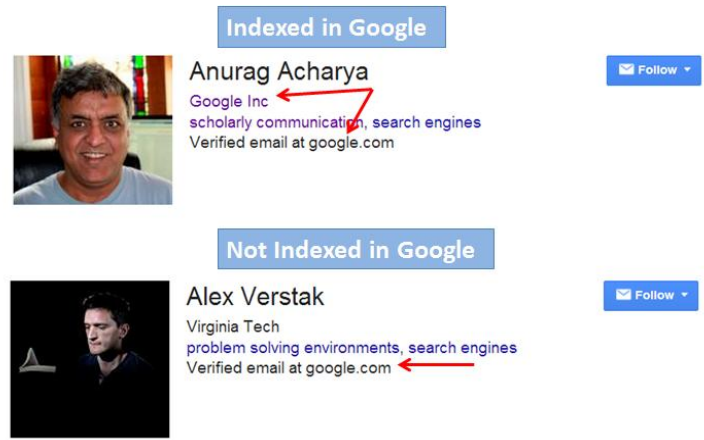

Fig.7. Identification of standardize institutions

In the case of *Anurag's* profile, both the institutional affiliation and the email do correspond with the company *Google*. Conversely, since the affiliation declared by *Verstak* (*Virginia Tech*) and his verified email (*google.com*) don't match, he hasn't been assigned to either of them.

**a) Google case studio**

Let's delve a little deeper into the *Google* case as an standardized institutional affiliation within *Google Scholar Citations*. We have identified a total of 1,043 scholars correctly ascribed on the main unified entry "*Google Inc*" (Figure 8).

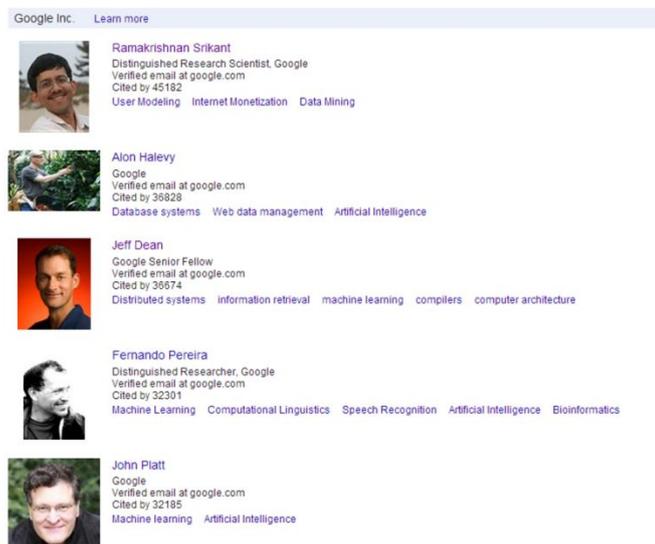

Fig.8. Scholars affiliated to *Google Inc.* in *Google Scholar Citations*

However, we find up to 1,116 profiles with an email linked to *Google* (*google.com*) but not declaring any professional affiliation with the company (Figure 9):

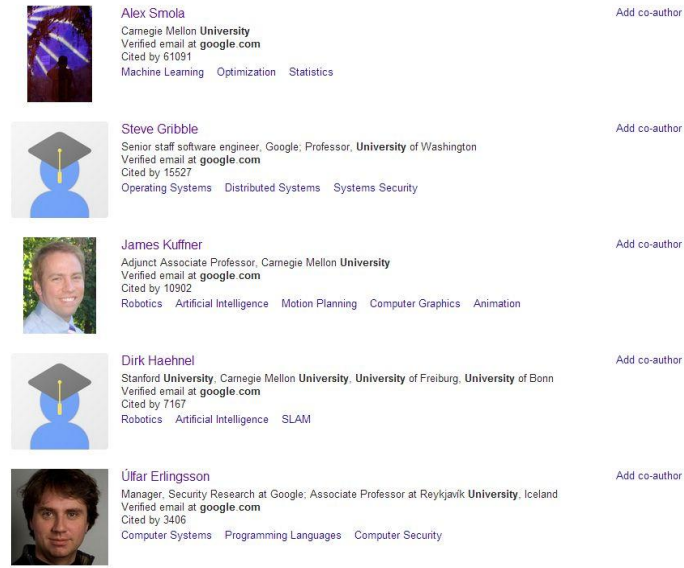

**Fig.9. Scholars not affiliated to *Google Inc.*(I): affiliation inconsistency**

Finally, we locate 465 profiles in which *Google* is declared as institutional affiliation although they do not indicate an institutional email from the company (Figure 10).

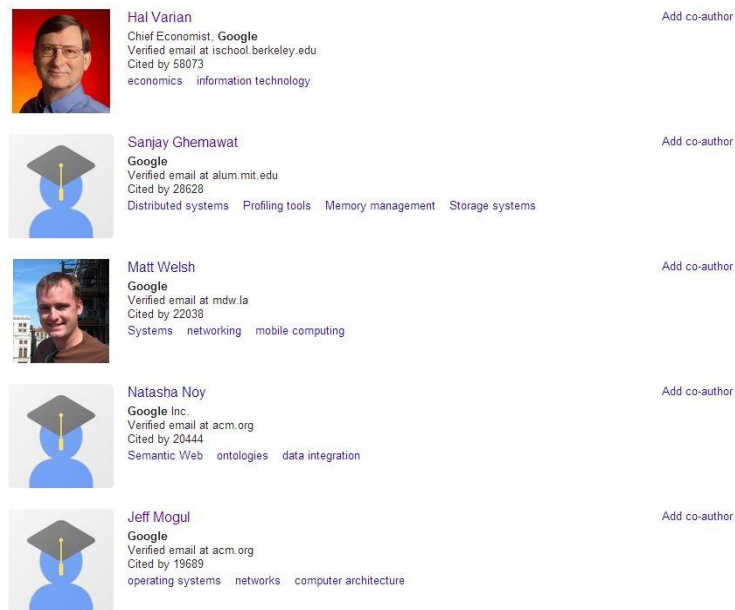

**Fig.10. Scholars not affiliated to *Google Inc.* (II): email inconsistency**

If we assume that all scholars who have mentioned either institutional affiliation to *Google* or an account of the institutional email are probably linked (currently or in the past) to this institution, approximately 538 scholars have been excluded, that is, almost a third of the total count considering all queries.

### b) Spanish universities case studio

We performed the same kind of searches with the Spanish universities finding similar results. Table I shows for each university the number of profiles gathered under the controlled institutional affiliation. Additionally we supply the number of profiles recovered under a query by email web domain (for example, ugr.es). Finally, the number of profiles under both the web domain and the institutional affiliation keyword queries (for example "*Universidad de Granada*)" are provided as well.

**Table I. Author profiles affiliated to Spanish universities in Google Scholar Citations**

| UNIVERSITY | Institutional link | Email | Email or affiliation |
|---|---|---|---|
| Universidad de Granada | **1,136** | 1,438 | 1,523 |
| Universidad Complutense de Madrid | **860** | 1,120 | 1,195 |
| Universitat Politècnica de Catalunya | **726** | 887 | 952 |
| Universitat Pompeu Fabra | **722** | 796 | 868 |
| Universitat de Barcelona | **674** | 834 | 1,042 |
| Universitat Autónoma de Barcelona | **636** | 787 | 886 |
| Universitat Politècnica de Valencia | **628** | 763 | 784 |
| Universidad de Sevilla | **613** | 825 | 891 |
| Universitat de Valencia | **592** | 780 | 846 |
| Universidad Politécnica de Madrid | **561** | 690 | 731 |
| Universidad Autónoma de Madrid | **522** | 656 | 705 |
| Universidad del País Vasco | **492** | 602 | 656 |
| Universidad Carlos III de Madrid | **469** | 549 | 598 |
| Universidad de Zaragoza | **453** | 583 | 621 |
| Universidad de Málaga | **438** | 543 | 573 |
| Universidad de Santiago de Compostela | **394** | 493 | 533 |
| Universidad de Alicante | **356** | 435 | 476 |
| Universidad de Murcia | **354** | 460 | 488 |
| Universidad Nacional de Educación a Distancia | **348** | 418 | 497 |
| Universidad de Cádiz | **321** | 402 | 420 |
| Universidad de Salamanca | **315** | 407 | 437 |
| Universidad de Castilla-La Mancha | **305** | 373 | 396 |
| Universidad de Oviedo | **302** | 365 | 404 |
| Universidad de Las Palmas de Gran Canaria | **286** | 349 | 365 |
| Universitat Jaume I | **281** | 330 | 339 |
| Universidad de Valladolid | **276** | 352 | 375 |
| Universidad de La Coruña | **273** | 327 | 332 |
| Universidad de Vigo | **258** | 317 | 338 |
| Universitat Rovira i Virgili | **254** | 290 | 328 |
| Universidad de La Laguna | **253** | 181 | 344 |
| Universidad de Extremadura | **244** | 306 | 319 |
| Universidad Rey Juan Carlos | **222** | 284 | 315 |
| Universitat de Girona | **217** | 253 | 290 |
| Universidad de las Islas Baleares | **208** | 231 | 264 |
| Universidad Miguel Hernández de Elche | **198** | 252 | 266 |
| Universidad de Navarra | **195** | 237 | 286 |
| Universidad de Alcalá | **190** | 240 | 265 |
| Universidad de Córdoba | **179** | 232 | 253 |
| Universidad de Jaén | **178** | 214 | 222 |
| Universidad de Cantabria | **173** | 220 | 229 |
| Universidad Pablo de Olavide | **169** | 217 | 231 |
| Universidad de León | **135** | 179 | 183 |
| Universitat Oberta de Catalunya | **133** | 168 | 190 |
| Universidad de Almería | **121** | 151 | 158 |
| Universitat de Lleida | **117** | 153 | 181 |
| Universidad de Huelva | **116** | 156 | 173 |
| Universidad de Deusto | **102** | 123 | 127 |
| Universidad Politécnica de Cartagena | **90** | 110 | 116 |
| Universidad Pública de Navarra | **90** | 108 | 109 |
| Universidad CEU Cardenal Herrera | **81** | 117 | 128 |
| Universidad CEU San Pablo | **81** | 117 | 124 |
| Universidad de Burgos | **62** | 80 | 87 |
| Universidad Ramon Llull | **59** | 66 | 66 |
| Universidad Católica San Antonio | **56** | 91 | 99 |
| Universidad Europea de Madrid | **54** | 61 | 68 |
| Universidad Internacional de La Rioja | **51** | 63 | 73 |
| Universidad de La Rioja | **45** | 51 | 56 |
| Universidad de Vich | **35** | 44 | 48 |
| Universidad Pontificia Comillas | **30** | 75 | 85 |
| Universidad IE | **30** | 41 | 42 |
| Universidad Loyola Andalucía | **30** | 37 | 49 |
| Universidad Católica de Valencia San Vicente Mártir | **27** | 38 | 39 |
| Universidad Camilo José Cela | **24** | 30 | 38 |
| Universidad a Distancia de Madrid | **22** | 26 | 26 |
| Universidad Internacional de Cataluña | **19** | 31 | 36 |
| Universidad San Jorge | **17** | 20 | 21 |
| Universidad de Mondragón | **15** | 25 | 26 |
| Universidad Pontificia de Salamanca | **13** | 17 | 24 |
| Universidad Francisco de Vitoria | **12** | 15 | 19 |
| Universidad Internacional Isabel I de Castilla | **0** | 5 | 5 |
| Universidad Antonio de Nebrija | **0** | 21 | 25 |
| Universidad Abad Oliva CEU | **0** | 6 | 8 |
| Universidad Europea Miguel de Cervantes | **0** | 6 | 7 |
| Universidad Católica Santa Teresa de Jesús de Ávila | **0** | 3 | 3 |
| Universidad Internacional Valenciana | **0** | 3 | 3 |
| Universidad Eclesiástica San Dámaso | **0** | 2 | 4 |
| Universidad Europea de Canarias | **0** | 2 | 3 |
| Universidad Europea de Valencia | **0** | 2 | 3 |
| Universidad Europea del Atlántico | **0** | 2 | 2 |
| Universidad Internacional de Andalucía | **0** | 1 | 5 |
| Universidad Alfonso X el Sabio | **0** | 1 | 4 |
| Universidad Internacional Menéndez Pelayo | **0** | 0 | 0 |
| **TOTAL** | **17,938** | **22,285** | **24,346** |

The data contained in Table I tell us that *Google Scholar Citations* collects 73.7% of all the scholars that could potentially have been linked to the institution by having an email or affiliation of the institution, and 80.5% of those with an official email from the institution. There are therefore a very significant number of scholars without proper institutional linkage.

Additionally we can observe up to 14 institutions (all of them private universities) with 0 authors included nor any standardized affiliation created (in some cases no valid emails are provided or the institutional name is not correct). We should point out the unexpected percentage obtained for the *University of Barcelona* (only 64.7% of authors included of the 1,042 potential authors with public profile created).

The conclusion after these brief case studies is that the standardized institutional affiliation link still currently has a number of shortcomings and pitfalls.

### *Methodological errors: does the identification algorithm operate properly?*

Because of the situation previously described - the most common - we can find examples (see Figure 11) where the algorithm does not work properly even though the profile apparently meets the requirements set by *Google Scholar Citations* (concordance between the affiliation name and the verified email).

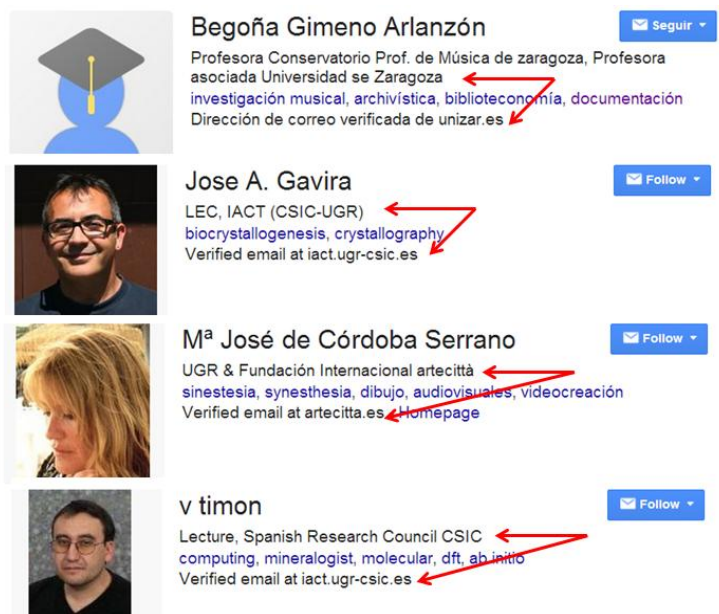

**Fig.11. Errors in the institutional affiliation linkage**

Next we identify and categorize some of these errors:

*a) Wrong uniform title*

In some cases *Google Scholar Citations* has not chosen as uniform title the official title of the institution, but a different one. For example, the *Catholic University San Antonio of Murcia* is identified as *Catholic University of Murcia* (see Figure 12). In this case *Google Scholar Citations* is probably influenced by the information located in *Wikipedia*, used to construct its knowledge graph (Figure 12 right).

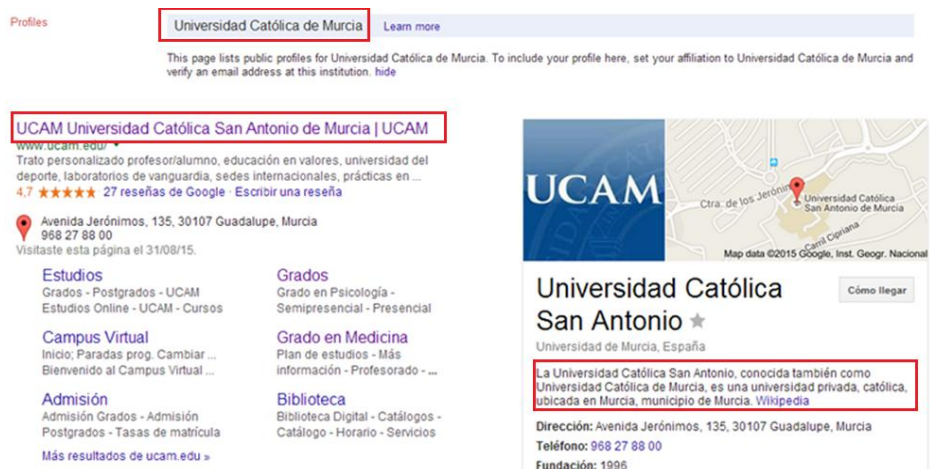

**Fig.12. Selection of erroneous uniform title for institutions**

b) *Misspelled affiliation*

For example, in the first profile showed in Figure 11, the affiliation name is "*Universidad se Zaragoza*". In this case, the system does not recognize the correct name of the affiliation. This reveals that the system apparently currently works with a controlled vocabulary of institutions.

c) *Name disambiguation problems*

The system doesn't use topographic qualifiers for discriminating different universities with the same name. This is the case of the *Universidad del Valle*, a name used by four different institutions in four different countries: Colombia, Guatemala, Mexico, and Bolivia. In this case, the tool is referring to the one based in Colombia (Figure 13).

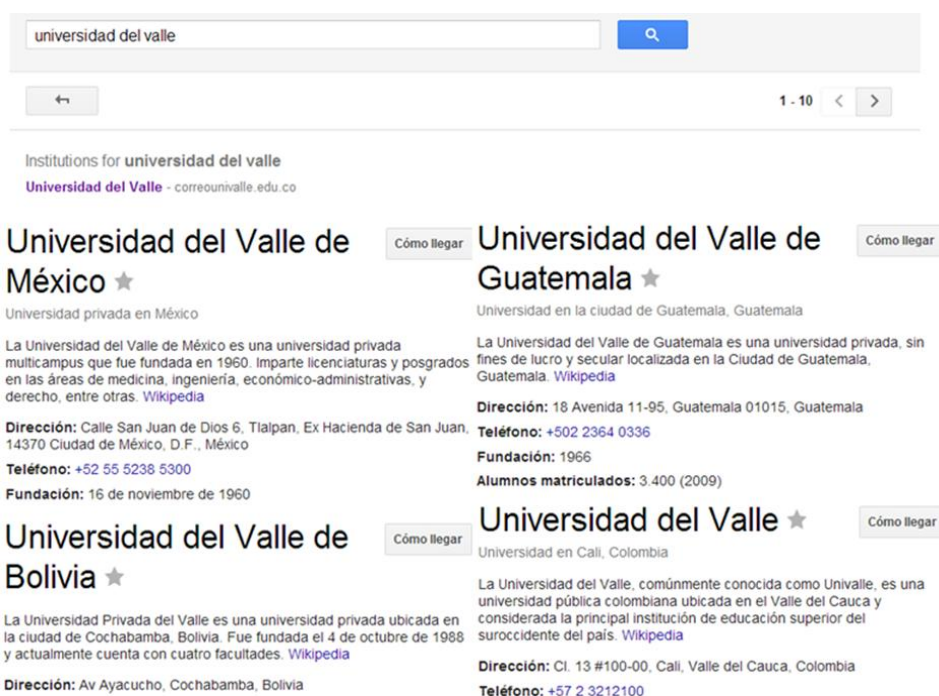

Fig.13. Institution names disambiguation

d) *Acronym disambiguation problems*

In some cases the identification of affiliations is not done correctly. For example, *Universidad CEU San Pablo, Universidad CEU Cardenal Herrera, and Universitat Abat Oliva CEU* are juridically independent universities, although they all belong to the same entity: *Centro de Estudios Universitarios – Center for University Studies* (CEU).

When *Universidad Cardenal Herrera* is identified as an affiliation (Figure 14a), the link points to *Universidad CEU San Pablo* (Figure 14b). This error may be attributed to the great variety of cases caused by the very different legislations applicable in each country.

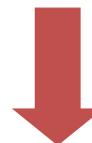

**Fig.14. Combination of different affiliations**

e) *Usage of more than one official academic web domain*

Some universities use different academic web domains for search engine optimization purposes (which in the case of universities is considered a bad practice)[6]. In these cases, the tool does not work accurately, as only one web domain is considered as valid (that is, web domain variants are not functioning as institution name variants). Following with the previous example, for *Universidad CEU Cardenal Herrera* we can find two different web domains (uch.ceu.es and uchceu.es). In Figure 15 we can find examples of two variants of verified email addresses for the same university. In this case it seems that "uch.ceu.es" is being considered as the valid web domain, so the affiliation in the second profile is not correctly linked.

**Fig.15. Institutions with more than one valid web domain**

*f) Lack of precision of the search box*

Additional problems have also been detected in the search box, which does not always correctly fulfill the task of identifying all the institutions that contain a particular keyword. Therefore, if the term "*Madrid*" is queried, the tool only returns two institutions (Figure 16), when actually there are six institutions that contain this word in the official name of the institution, and which have been clearly identified by *Google Scholar Citations* under other specific queries, as shown in Figure 16 (down).

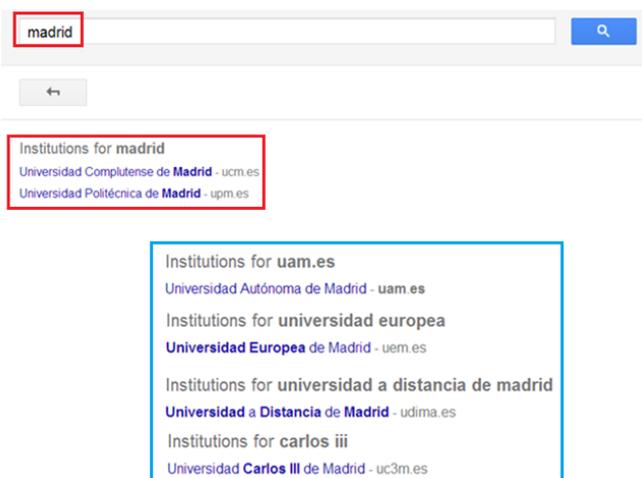

**Fig.16. Recovering institutions in the Google Scholar Citation search box I**

More problematic is the case of the search for the term "*Católica*" (Catholic), which retrieves only one institution, where *Google Scholar Citations* has identified at least 14 institutions using that term (Figure 17):

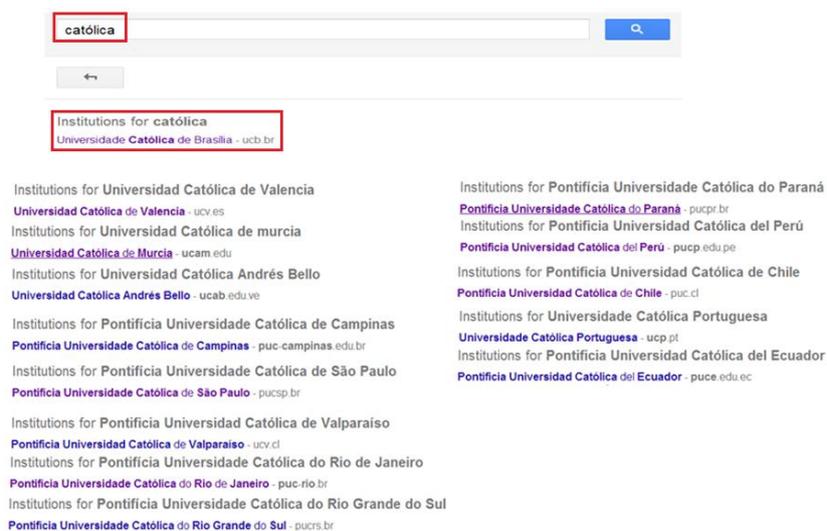

**Fig.17. Recovering institutions in the Google Scholar Citation search box II**

Interestingly, the search seems to work better with Anglo-Saxon institutions (Figure 18). If we query "Maryland", a comprehensive list of institutions with this word in their name is obtained. However, if you are searching "Loyola", Anglo-Saxon institutions are identified, but Spanish institutions containing this

keyword, for example *Universidad Loyola de Andalucía* (*Loyola University of Andalusia*), are not included in this list, even though we know the institution is already controlled and identified in the system, because it can be found with a more specific query.

**Fig.18. Recovering institutions in the *Google Scholar Citation*s search box III**

*g) Complex institutions*

We encountered various problems having to do with complex institutions. This is the case of the *Spanish National Research Council* (CSIC) in Spain. Although the standard term in *Google Scholar Citations* to this institution is "CSIC", mixed centers (ventures between universities and research institutions belonging to CSIC) are not linked correctly.

For example, in the case of our colleague *Ismael Rafols* (CSIC-UPV), the system links the author to the UPV (Figure 19), instead to CSIC, which strictly speaking would be his correct affiliation.

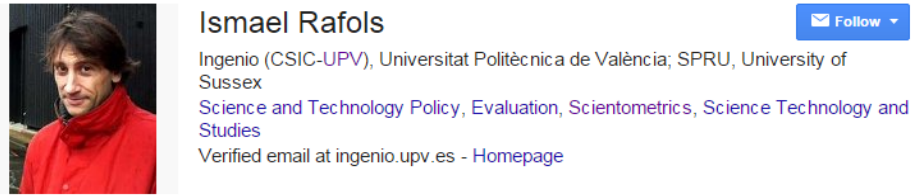

**Fig.19. Complex institutions' affiliation in Google Scholar Citation**

Similarly, authors who have introduced their affiliation membership by adding the specific research institute (in the form Research institution – CSIC or institution / CSIC), do not appear linked to any institution, and therefore are not integrated in the CSIC profile.

*h) Multiple affiliations*

Lastly, authors with more than one affiliation represent another group of profiles which have not been correctly affiliated. This is the case, for example, of Matthew Fujita (*University of Texas at Arlington*, *University of California Berkeley, Harvard and Oxford*, nothing more, nothing less…). In Figure 20 we can observe how only one institution appears with the affiliation hyperlink (in this case *Texas at Arlington*). If we carried out a keyword search by the query "*Harvard*", this author would appear in the list but will not be included within the *Harvard University* institution profile. The reason behind this limitation is that only one email is provided (*uta.edu*).

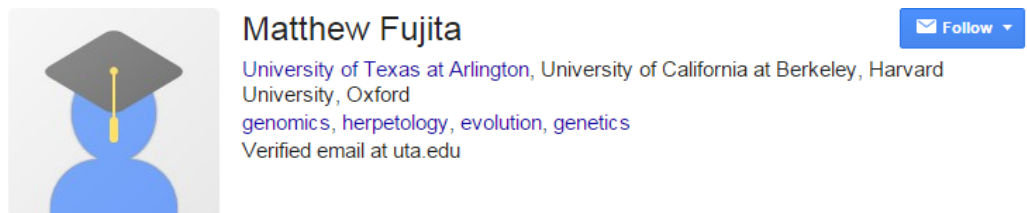

**Fig.20. Authors with multiple institutional affiliations**

*i) Internal affiliations*

Following with the previous example, we also noticed that authors who include internal affiliations (departments, research institutes, Faculties, etc.), are not included. If we search by "*Harvard Medical School*", we can find authors that only provide this affiliation (Figure 21), for example Michael A. Moskowits. No affiliation is linked in these cases.

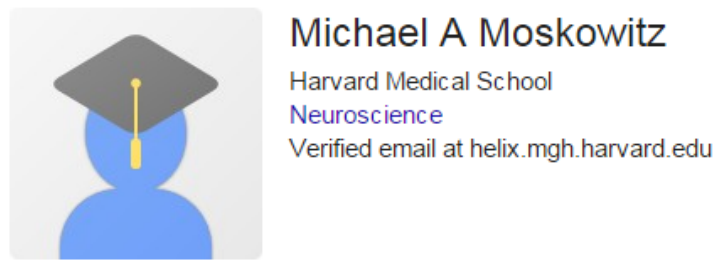

**Fig.21. Authors with internal institutional affiliations**

## 6. FINAL RECOMMENDATIONS AND SUGGESTIONS

We should not be surprised about inconsistencies and errors made by Google Scholar because the task of authority control is devilishly complex. However, these errors are the natural consequence of a product conceived with the "laissez faire laissez passer" principle in mind. We already warned about this in our post commenting on the <u>changes introduced in 2014 in Google Scholar Profiles</u>[7].

"Total freedom is given to authors in order to set out the profile to their own taste and benefit, from their personal, professional and thematic identification to the possibility of linking bibliographic production they deem appropriate or even selecting the mode of profile updates (automatic or manual). The author can add, edit and delete bibliographic records as he/she pleases. *Google Scholar Citations* is unaware of the possible truthfulness and accuracy of the data and leaves to the author the full responsibility for what is shown.

This wide autonomy granted to the author reflects Google's philosophy: a set of user-oriented information; created by and for the user. A clever way not only to reduce the costs derived from treatment and information management but also to make the user work for the system". Nonetheless, a solution - as we can see - with its own risks if authority control is not provided at all.

Finally we deem it necessary to formulate a series of recommendations for *Google Scholar Citations* users and other actors involved in this product, all of them aimed at improving it.

### a) Google Scholar Citations

- Extending the user information form that has to be filled the profile is created, providing fields to store information about the institution and the country. These fields should incorporate suggestions about the standard name of the institution and the correct acronym.
- Allowing authors to include several institutions, which will be much useful for all those who change them throughout their academic life. For example, an open box for Past Institutions and years of affiliation.

Table II. Example of affiliation information to be filled in the profile

| Institution name | Acronym | Country | Date |
|---|---|---|---|
| Universidad de Granada | UGR | Spain | 1980-85 |
| Universitat Politècnica de València | UPV | Spain | 1986- |

In order to guide users to standardize institutions, *Google* can use a similar technique to the one it already uses in the general search engine, that is, to offer different standardized options available for selection as the user begins typing inside the text box. In the case of the country field, a dropdown selection box might be the easiest solution.

To avoid omissions (deliberate or not), the use of mandatory fields may be adopted. In this way, many problems are avoided at the same time the profile is created, which is the best option to develop an accurate and comprehensive authority control in an open system like this. All the same, *Google* has never focused (neither in the general search engine nor in the academic one) in such controlled-vocabulary approaches, but instead has usually relied on natural language searching. The combination of these two techniques may help feeding the system with more acurate information, therefore improving its precision.

**b) Authors**

We might advise the authors about the repercussions of not including the name of the institution they are affiliated with, as well as not using an e-mail account from that same institution to verify the profile, because if these requirements are not met they will be directly excluded from the university aggregation constructed by the system. Moreover, they must not only indicate the institution name, but they should use the standardized nomenclature.

The attention paid to this subject reflects that institutional names should not be translated but expressed in the original language (*Universidad de Jaén* instead of *University of Jaen*).

Although today it is a widespread practice to include the names of the centers in English to facilitate international visibility, in order to prevent the proliferation of variants this practice is advisable only when the center has an accepted and standardized name in this language.

*c) Institutions*

We must make an appeal to the institutions appearing on the *Google Scholar Citations* so that they don't ever consider elaborating evaluative products without first conducting a thorough search of all scholars that may be currently linked to their institution.

These institutions should also develop specific policies to encourage its staff to use its corporate name in a standardized way.

We do hope the profound changes experienced by *Google* this summer (now Alphabet), do not pose any limitation to Google Scholar so that it can keep growing and improving, not only as an academic search engine but as a valuable data source to bibliometrics.

# REFERENCES

1. Verstak, A. (2014, August 21). Fresh Look of Scholar Profiles [Blog Post]. In *Google Scholar Blog*. Retrieved from http://goo.gl/QQdifH

2. Aguillo, I. [isidroaguillo]. (2015, August 26). Google Scholar Citations add links to institution`s names (incl acronyms) in correct-built affiliations of profiles https://t.co/ellBYE3056. Retrieved from https://goo.gl/EhSUux

3. Authority Control. (n.d.). In *Wikipedia*. Retrieved September 13, 2015, from https://en.wikipedia.org/wiki/Authority_control

4. Virtual International Authority File. Retrieved on 13/09/2015 from https://viaf.org/

5. Delgado López-Cózar, E. (1997). Incidencia de la normalización de las revistas científicas en la transferencia y evaluación de la información científica. *Rev neurol*, 25(148): 1942-1946

6. Orduña-Malea, E. (2012). Graphic, multimedia, and blog content presence in the Spanish academic web-space. *Cybermetrics: International Journal of Scientometrics, Informetrics and Bibliometrics*, *16*(1).

7. Delgado López-Cózar, E. (2014, August 24). Google Scholar da un lavado de cara a sus Google Scholar Profiles=Google Scholar Citations [Blog Post]. In EC3noticias. Retrieved from http://goo.gl/z8U4K0

## ACKNOWLEDGEMENTS

Juan Manuel Ayllón has a FPI pre-doctoral scholarship for research (BES-2012-054980) funded by the Spanish Ministry of Economy and Competitiveness. Alberto Martín-Martín holds a fellowship for the training of university teachers (FPU2013/05863), funded by the Spanish Ministry of Education, Culture and Sport.